\documentclass[12pt]{article}

\usepackage{epsfig}

\def\bb{\mbox{\bf b}}
\def\bd{\mbox{\bf d}}
\def\be{\mbox{\bf e}}

\def\bs{\mbox{\bf s}}
\def\bt{\mbox{\bf t}}
\def\bc{\mbox{\bf c}}
\def\bu{\mbox{\bf u}}

\def\balpha{\mbox{\boldmath $\alpha$}}

\def\btau{\mbox{\boldmath $\tau$}}
\def\bnu{\mbox{\boldmath $\nu$}}
\def\bmu{\mbox{\boldmath $\mu$}}

 \def\tfrac#1#2{{\textstyle{{#1}\over{#2}}}}

\def\half{\tfrac{1}{2}}

\def\quart{\tfrac{1}{4}}

\def\Im{\mathop{\rm Im}\nolimits}

\def\tr{\mathop{\rm tr}\nolimits}
\def\Tr{\mathop{\rm Tr}\nolimits}

\usepackage{amssymb}

\begin{document}

\begin{titlepage}

\baselineskip 24pt

\begin{center}

{\Large {\bf $\delta_{CP}$ for leptons and a new take on CP physics 
with the FSM}}

\vspace{.5cm}

\baselineskip 14pt

{\large Jos\'e BORDES \footnote{Work supported in part by Spanish
    MINECO under grant FPA2017-84543-P and PROMETEO 2019-113 (Generalitat Valenciana). }}\\
jose.m.bordes\,@\,uv.es \\
{\it Departament Fisica Teorica and IFIC, Centro Mixto CSIC, Universitat de 
Valencia, Calle Dr. Moliner 50, E-46100 Burjassot (Valencia), 
Spain}\\
\vspace{.2cm}
{\large CHAN Hong-Mo}\\
hong-mo.chan\,@\,stfc.ac.uk \\
{\it Rutherford Appleton Laboratory,\\
  Chilton, Didcot, Oxon, OX11 0QX, United Kingdom}\\
\vspace{.2cm}
{\large TSOU Sheung Tsun}\\
tsou\,@\,maths.ox.ac.uk\\
{\it Mathematical Institute, University of Oxford,\\
Radcliffe Observatory Quarter, Woodstock Road, \\
Oxford, OX2 6GG, United Kingdom}

\end{center}

\vspace{.3cm}

\begin{abstract}

A bonus of the framed standard model (FSM), constructed initially to 
explain the mass and mixing patterns of quarks and leptons, is a
solution (without axions) of the strong CP problem by cancelling the 
theta-angle term $\theta_I \Tr (H^{\mu \nu}H^*_{\mu \nu})$ in colour
\footnote{We use $F_{\mu \nu}, G_{\mu \nu}$, and $H_{\mu \nu}$ here 
to denote respectively the $u(1)$, flavour $su(2)$ and colour $su(3)$ 
gauge fields.} by a chiral transformation on a quark zero mode which 
is inherent in FSM, and produces thereby a CP-violating phase in the 
CKM matrix similar in size to what is observed \cite{tfsm}.  Extending 
here to flavour, one finds that there are two terms proportional to 
$\Tr (G^{\mu \nu} G^*_{\mu \nu})$: (a) in the action from flavour 
instantons with unknown coefficient, say $\theta'_I$, (b) induced 
by the above FSM solution to the strong CP-problem with therefore 
known coefficient $\theta'_C$.  Both terms can be cancelled in the 
FSM by a chiral transformation on the lepton zero mode to give a 
Jarlskog invariant $J'$ in the PMNS matrix for leptons of order 
$10^{-2}$, as is hinted by experiment.  But if, as suggested in 
\cite{cpslash}, the term $\theta'_I$ is to be cancelled by a chiral 
transformation in the predicted hidden sector to solve the strong CP 
problem therein, leaving only the term $\theta'_C$ to be cancelled by 
the chiral transformation on leptons, then the following prediction 
results: $J' \sim - 0.012$ ($\delta'_{CP} \sim (1.11) \pi$) which is 
(i) of the right order, (ii) of the right sign, (iii) in the range 
favoured by present experiment.  Together with the earlier result for 
quarks, this offers an attractive unified treatment of all known CP 
physics. 

\end{abstract}   

\end{titlepage}

\newpage

\section{Preamble}

One notable result of the framed standard model (FSM) \cite{fsmpop}, 
constructed initially to understand the mass and mixing patterns of 
quarks and leptons, is in offering a solution (without axions) to 
the strong CP problem by converting the theta-angle term in the QCD 
action into a CP-violating phase $\delta_{CP}$ \cite{KM} in the CKM matrix, 
thereby answering the question of origin for weak CP-violation as 
well.

This result comes about as follows. The mass matrix for quarks and 
leptons in FSM is of the following form:
\begin{equation}
m = m_T \balpha \balpha^\dagger,
\label{mfact}
\end{equation}
where $\balpha$, a real unit vector in 3-dimensional generation 
space, is the same for all quarks and leptons, and only the scalar 
factor $m_T$ depends on the fermion type.  Under renormalization, 
$\balpha$ changes with scale (rotates) in a prescribed manner 
governed by renormalization group equations (RGE), and gives rise 
to an intricate mass and mixing pattern for quarks and leptons.  
By adjusting the 7 parameters left unspecified by the RGE, the fit 
to data shown in Table 2 of \cite{tfsm} is obtained, for the 
masses and mixing angles of quarks and leptons, mostly to within 
the present experimental accuracy.  As to the very relevant 
question how the rank-one mass matrix (\ref{mfact}) gives rise, 
when $\balpha$ rotates, to nonzero masses for all quarks and 
leptons, the answer is too long to be summarized here 
\footnote{The point is that for 
a rotating mass matrix such as (\ref{mfact}), which has both its 
eigenvalues and eigenvectors changing with scale, the masses and 
state vectors of the physical states are not given just by the 
eigenvalues and eigenvectors at any one scale.  Hence, a zero 
eigenvalue at any scale need not imply an actual particle with 
vanishing physical mass.  However, for the CP problems 
to be dealt with in this paper, what matter are indeed the 
eigenvectors with zero eigenvalue of the mass matrix taken at 
some specified scale, which we shall also refer to sometimes as 
states with no mass terms when appropriate, and sometimes just 
as ``zero modes'' for short.} but can be 
found in, for example, \cite{tfsm}.

Note that (\ref{mfact}), being of rank one at every scale $\mu$, 
has a zero eigenvalue in the direction of the binormal $\bnu$, 
orthogonal to both $\balpha$ and the tangent $\btau$ to the curve 
traced out by $\balpha$.\footnote{The mass matrix (\ref{mfact}) 
has of course also a zero eigenvalue in the direction of the 
tangent $\btau$, but the chiral transformation that we wish to do 
next in the $\bnu$ direction, if performed instead on $\btau$, 
would make the vector $\balpha(\mu + \delta \mu) = \balpha + 
\delta \mu \btau$ complex at the neighbouring scale $\mu + \delta 
\mu$, contrary to the RGE governing the $\mu$-dependence, which 
keeps $\balpha$ real \cite{tfsm} and, being perturbatively 
derived, is unaffected by the theta-angle term.}  Now it is well 
known that once given quarks with zero mass eigenvalue, then the 
theta-angle term,\footnote{Following Weinberg \cite{Weinbergqft}, 
we shall call all such terms theta-angle terms, whether they are 
proportional to $\Tr(H^{\mu \nu} H^*_{\mu \nu})$ in colour or to 
$\Tr(G^{\mu \nu} G^*_{\mu \nu})$ in flavour, or whether they arise 
in the action from instantons or appear in the measure of Feynman 
path integrals as a result of chiral transformations on fermion 
fields, and also whatever their coefficients may be, whether 
$\theta$ or $\alpha$ or some other number.} say from instantons,
in the Lagrangian density of the QCD action:     
\begin{equation}
- \frac{(\theta_I)}{16 \pi^2} \tr_C (H^{\mu \nu} H^*_{\mu \nu})
\label{thetatermC}
\end{equation}
(where $\tr_C$ is the trace over colour indices) which is 
responsible for the strong CP problem can be cancelled in 
CP-violating effects by a chiral transformation on those quark 
fields.   

Explicitly, under the chiral transformation:
\begin{equation}
\psi^f_0 \rightarrow (\exp i \alpha^f \gamma_5) \psi^f_0
\label{chiraltrans}
\end{equation}
on the quark field $\psi^f_0$ with zero mass eigenvalue, the measure 
in Feynman path integrals is known \cite{Weinbergqft,Fujikawa} to 
acquire an extra term in the exponent:
\begin{equation}
- \frac{\left( 2 \sum_f \alpha^f \right)} {16 \pi^2} 
   \tr_C (H^{\mu \nu} H^*_{\mu \nu}) ,
\label{extraterm}
\end{equation}
of the same form as the theta-angle term (\ref{thetatermC}) apart 
from the coefficient.  Hence, in particular, if one chooses: 
\begin{equation}
\left( \sum_f \alpha^f \right) = - \half \theta_I,
\label{sumf}
\end{equation}
then the CP-violating theta-angle term (\ref{thetatermC}) will be 
cancelled by (\ref{extraterm}) in all Feynman path integrals for 
physical quantities.  (Details of the notation used here can be 
found in Appendix A, where also some assertions seemingly deduced 
heuristically in the text will be referenced in detail or more 
formally derived.)  

Physically, what this means is the following.  For a massive 
fermion represented by a Dirac 4-component spinor $\psi$, its 
mass term, $m \bar{\psi} \psi$, fixes the relative phase between 
its CP-conjugate left- and right-handed components.  A change in 
that relative phase, as effected by a chiral transformation on 
$\psi$, will make the mass parameter complex, hence artificially 
CP-violating and unacceptable.  (See Appendix B for more details.)  
However, a massless fermion, with no mass term to keep real, is 
under no such constraint.  In other words, CP is defined for a 
massless $\psi_0$ only up to this relative phase between its left- 
and right-handed components.  This applies also to any state with 
a zero eigenvalue of the mass matrix in the multi-generation case.
Now, the mass matrix (\ref{mfact}) does have quark states with 
zero eigenvalue in the direction $\bnu$.  Hence one is free to 
choose, in the definition of CP for these states, the relative 
phase between their left- and right-handed components so as to 
cancel the theta-angle term (\ref{thetatermC}), as is done by 
the above chiral transformation (\ref{chiraltrans}), and so 
``solves the strong CP-problem''.     

This can be done at every scale $\mu$.  However, $\bnu$ is orthogonal 
to $\balpha$ and $\balpha$ rotates with changing $\mu$.  Thus the 
chiral transformation needed to cancel the theta-angle term would 
differ from $\mu$ to $\mu$ also.  It follows then that preserving 
CP at one $\mu$ by this means does not guarantee the same at 
another $\mu$.  Now the elements of the CKM matrix do involve 
quantities evaluated at different values of $\mu$.  Hence, some 
degree of CP-violation would remain in the CKM matrix by virtue of 
the changing direction of $\bnu$ (or the rotation of $\balpha$) 
under scale changes.  And this, in FSM, is what gives rise to the 
Kobayashi-Maskawa (KM) CP-violating phase.  The details were worked out in 
\cite{strongCP,atof2cps} where it was further noted that a $\theta_I$ of 
order unity in (\ref{thetatermC}) gives automatically a Jarlskog 
invariant \cite{Jarlskog} of the order of magnitude $10^{-5}$ seen 
in experiment.  These considerations have been taken fully into 
account in the fit \cite{tfsm} and are seen to affect materially 
the quality of the fit. 

In summary, as far as quarks are concerned, the FSM seemed to 
have done two things:
\begin{itemize}
\item It has identified what were at first sight two separate CP 
problems: namely, the strong CP problem embodied in the 
theta-angle term in the QCD action and the weak CP problem 
embodied in the CP-violating phase of the CKM matrix.
\item It has exorcised the potentially huge CP-violation effects 
from the theta-angle term which are much beyond what experiment 
allows, and reduced it to the size range actually seen in 
experiment, namely just as a phase $\delta_{CP}$ in the CKM matrix.
\end{itemize}
Note that although CP-violation via the $\delta_{CP}$ phase is sometimes 
referred to as a weak (flavour) effect because the CKM matrix 
appears in the standard model (SM) in the electroweak coupling, in the FSM 
framework here, it is generated entirely within the colour 
sector via colour framon loops, and it is of a limited size 
as seen in experiment only because it is a loop effect and 
therefore bound to be perturbatively small.
 
The two results were obtained in the FSM by virtue of the two 
particular properties that it possesses:
\begin{itemize}
\item {\bf [P1]} The fermion mass matrix is such that there is 
one direction in generation space where quark states have a zero 
mass eigenvalue so that a chiral transformation can be made on 
them without making any mass complex.
\item {\bf [P2]} This direction in generation space depends on 
scale and it is its change with scale which gets translated 
into the $\delta_{CP}$ phase.
\end{itemize} 
Indeed, qualitatively similar results to these were obtained 
in phenomenological models \cite{atof2cps,r2m2,bjpaper} starting 
with {\bf [P1]} and {\bf [P2]} as ans\"atze before the 
FSM was formulated.  But the FSM has made these erstwhile 
ans\"atze into consequences and given thereby more precise 
results with fewer adjustable parameters.  This remark will be 
useful later in assessing the actual relevance of the FSM to 
new results obtained hereafter.  

\section{CP violation in leptons from a theta-angle term}

The above considerations from QCD do not concern leptons since 
these do not carry colour, but given that a theta-angle term, 
say:
\begin{equation}
- \frac{(\theta'_I)}{16 \pi^2} \tr_F (G^{\mu \nu} G^*_{\mu \nu})
\label{thetatermF}
\end{equation}
should appear also in the flavour theory on the same topological 
grounds, and that the properties {\bf [P1]} and {\bf [P2]} in 
FSM hold also for leptons, our first instinct was that similar 
arguments should apply to the theta-angle term (\ref{thetatermF}) 
for the flavour theory, reducing it to a CP-violating phase
$\delta'_{CP}$ in the 
PMNS matrix for leptons as well.  But we were dissuaded from 
doing so in \cite{tfsm} by a suggestion seen in the literature 
that we then accepted, namely:
\begin{itemize}
\item {\bf [S]} That the term (\ref{thetatermF}) in the flavour 
theory can be trivially transformed away even when all quarks 
and leptons are massive, and should thus give no physical effect.
\end{itemize}
However, now that there is increasing experimental evidence for 
a non-zero $\delta'_{CP}$ for leptons \cite{pdg,T2K}, there is the 
incentive for us to examine the problem anew.  We have done so, 
as is reported in Appendix B, and come to the conclusion that 
the arguments we have seen do not in fact lead to {\bf [S]}.  
Though of interest by itself, this analysis is not essential in 
the context of the present article, the main purport of which is 
to show that the presence of a theta-angle term in flavour such 
as (\ref{thetatermF}) will lead in the FSM to a non-vanishing 
CP-violating phase $\delta'_{CP}$ in the PMNS matrix for 
leptons.  Since $\delta'_{CP}$ is a physical quantity which is 
being measured---in fact a recent result from T2K \cite{T2K} 
has given at $3 \sigma$ level a CP violating phase for the PMNS 
matrix---this result, if confirmed, will already serve as a 
counter-example to {\bf [S]} and invalidate it, without recourse 
to the analysis given in Appendix B, to which it is therefore  
relegated.

As in the colour theory, a theta-angle term in flavour such as 
(\ref{thetatermF}) can potentially lead to huge CP-violations, 
such as baryon number $B$ and lepton number $L$ non-conservation, 
way beyond what is seen in experiment.  But such a disaster is 
avoidable in the FSM by the fact that there are at every scale 
$\mu$ quark and lepton states with zero mass eigenvalues whose 
CP properties are not {\it a priori} defined, on which states 
therefore appropriate chiral transformations can be performed to 
generate extra theta-angle terms from the measure of Feynman 
integrals to cancel the original ones.

There is, however, one important difference between the flavour 
theory and the colour theory.  Only quarks carry colour, so that 
chiral transformations only on quarks, but not on leptons, would 
generate theta-angle terms in colour from the measure of Feynman 
integrals.  On the other hand, both quarks and leptons carry 
flavour so that chiral transformations on either will generate 
theta-angle terms in flavour theory.  That this is so is shown 
more formally in Appendix A. 

Suppose then we start anew from an action with a theta-angle term
(\ref{thetatermC}) coming, say, from instantons in the colour 
theory, plus a theta-angle term (\ref{thetatermF}) coming from 
instantons in the flavour theory.  To avoid unacceptably large 
CP-violations in the colour theory, we follow what was done in for 
example \cite{tfsm} as described in the Preamble, and effect the 
chiral transformations (\ref{chiraltrans}) with $\sum_f \alpha^f 
= - \theta_I/2$ on the massless quark states along the direction 
$\bnu$.  This generates from the measure of Feynman integrals the 
extra term (\ref{extraterm}), so as to cancel the term in the 
Lagrangian density (\ref{thetatermC}) of the action and remove the 
danger of unacceptably large CP-violations in the colour sector.  
However, since quarks are flavour doublets, the transformation 
(\ref{chiraltrans}) will generate from the measure of Feynman 
integrals also a term, say: 
\begin{equation}
- \frac{(\theta'_C)}{16 \pi^2} \tr_F (G^{\mu \nu} G^*_{\mu \nu}).
\label{extratermCi}
\end{equation}
What value this $\theta'_C$ will take can be worked out from the 
details of the fit reported in \cite{tfsm}, as we shall do later.  
For the moment, we note only that we have now, in effect, two 
theta-angle terms to deal with in the flavour sector, namely both 
this new one (\ref{extratermCi}) with coefficient $\theta'_C$ 
induced by the solution of the strong CP problem in QCD, and the 
original one (\ref{thetatermF}) with coefficient $\theta'_I$ from 
instantons in the action.

Fortunately in the FSM there is also at every scale, in the $\bnu$ 
direction in generation space, some lepton states with zero mass 
eigenvalue, the CP phases of which are {\it a priori} undefined 
and on which chiral transformations are allowed.  By choosing 
appropriately the chiral transformation (\ref{chiraltrans'}) on 
leptons in the $\bnu$ direction: 
\begin{equation}
\psi_0^{f'} \rightarrow (\exp i {\alpha'}^{f'} \gamma_5) \psi_0^{f'},
\label{chiraltrans'}
\end{equation}
that is explicitly:
\begin{equation}
\sum_{f'} {\alpha'}^{f'} =  \alpha' = - (\theta'_I + \theta'_C),
\label{alpha'}
\end{equation}
one can remove completely the theta-angle term for flavour as 
well from all Feynman path integrals.  Note that equation 
(\ref{alpha'}) differs from  (\ref{extraterm}) by a factor
2 because only the left-handed lepton is coupled to the flavour 
field.  Interestingly, there is just enough freedom to remove both 
the theta-angle terms in colour and flavour and the solution is 
unique.

This can be done at every scale, but since $\bnu$ rotates with 
scale, the mechanism described in the Preamble will give rise to 
CP-violating phases in the mixing matrices, that is, both in the CKM 
matrix for quarks and in the PMNS matrix for leptons. 

In the colour case, it was already shown by the fit in \cite{tfsm} 
that a theta-angle term with $\theta_I$ of order unity will lead to 
a Jarlskog invariant $J$ of the order ($J \sim 10^{-5}$) seen in 
experiment \cite{pdg}.  What about the flavour case for leptons?
It turn out that with the additional information obtained from 
\cite{tfsm}, the parallel application to flavour is made easier 
and can be carried further, indeed nearly to the point of yielding 
a direct estimate for $J'$ (or $\delta'_{CP}$) for leptons, as we 
shall now demonstrate.

We begin with the clarification of some ambiguities and correction 
of an error in \cite{tfsm} which were of no practical import there
when our interest was only in the order of magnitude of $J$, but 
have now become relevant when our interest includes also the signs 
and actual values of both $J$ and $J'$.  

First, we need to clarify an ambiguity in notation.  
In equation (17) in \cite{tfsm}, the CKM matrix element say $V_{ud}$ was 
given as $\tilde{\bu} \cdot \tilde{\bd}$ to represent the scalar 
product between the two complex vectors without it being specified 
whether it meant $\tilde{\bu}^\dagger \tilde{\bd}$ or its complex 
conjugate $\tilde{\bd}^\dagger \tilde{\bu}$.  This caused no problem 
there as it dealt with only absolute values of CKM matrix 
elements and of $J$.  However, now that we are interested also in 
the sign of $J$, this ambiguity has to be removed.  Henceforth, in 
conformity with standard usage, the CKM matrix should read instead, 
 unambiguously as:
\begin{equation}
V_{CKM} = \left( \begin{array}{ccc}
\tilde{\bd}^\dagger \tilde{\bu} & \tilde{\bs}^\dagger \tilde{\bu} & 
   \tilde{\bb}^\dagger \tilde{\bu}  \\ 
\tilde{\bd}^\dagger \tilde{\bc} & \tilde{\bs}^\dagger \tilde{\bc} & 
   \tilde{\bb}^\dagger \tilde{\bc}  \\ 
\tilde{\bd}^\dagger \tilde{\bt} & \tilde{\bs}^\dagger \tilde{\bt} & 
   \tilde{\bb}^\dagger \tilde{\bt}
           \end{array} \right)
\label{VCKMnew}
\end{equation}

The above ambiguity means also an ambiguity in sign of 
$\theta_I$ and of $J$.  Changing the elements of $V_{CKM}$ to their 
complex conjugtes changes the sign of $J$, and in FSM also the signs 
of $\alpha_f$ since CKM matrix elements there become complex only by 
the insertion of phase factors $\exp \pm i \alpha_f$ in appropriate 
places, and hence the sign of $\theta_I$ which these $\alpha_f$ are 
supposed to cancel.  It has since been ascertained, in the fit of 
\cite{tfsm}, that to obtain $J$ positive, as wanted by experiment 
\cite{pdg}, then $\theta_I$ has also to be positive.
 
Lastly, we wish to correct an error in \cite{tfsm} so as to chime in 
with the spirit of the present article, although this will not affect 
any of the results here or materially those of \cite{tfsm}.  The fit 
there was obtained by a chiral transformation (\ref{chiraltrans}) on the quark 
zero mode of both the up- and down-types with $\alpha^f=-\alpha_0$,
where
\begin{equation}
\alpha_0 = 0.89,
\label{alpha0}
\end{equation}  
which we shall take as a sort of unit for ease of later presentation.
Since both the up- and down-quark zero modes were subjected there to 
(\ref{chiraltrans}), the formula (\ref{sumf}) should give then:
\begin{equation}
\theta_I = 4 \alpha_0 \sim  3.56,
\label{thetaI}
\end{equation}
that is, twice what was then thought.  (A more formal derivation of 
this factor of 2 will be given in Appendix A.)  This error does not 
affect results in this paper since (\ref{thetaI}) does not enter, 
while in \cite{tfsm} it was of no practical import since the concern 
there was only that a $\theta_I$ of order unity can give rise to $J$ 
of order $10^{-5}$, as experiment wanted. 
 
With these ambiguities and error cleared up, one can now proceed to 
details.  In \cite{tfsm}, one has obtained already, one believes, 
what should be fair approximations for the state vectors of all 
the 6 leptons.  These are given in the Appendix there, 
in the conveniently chosen coordinates there specified, as:
\begin{eqnarray}
  \btau^\dagger &=& (-0.90435, 0.31923, 0.28329)\nonumber\\
  \bmu^\dagger &=& (0.07434, -0.53580, 0.84107)\nonumber \\
  \be^\dagger &=& (0.42028, 0.78168, 0.46082) \nonumber\\
  \bnu_3^\dagger &=& (-0.62012, 0.07942, 0.78048) \nonumber \\
  \bnu_2^\dagger &=& (0.77494, -0.09291, 0.62517) \nonumber\\
  \bnu_1^\dagger &=& (0.12217, 0.99250, -0.00393)
\label{statevecs}
\end{eqnarray}

One has as well the rotation trajectory for $\balpha$ (Figure 3 
of \cite{tfsm}) from which one can deduce the Darboux triad
\footnote{When we consider a curve embedded on a given surface, 
as in our case of the trajectory of $\balpha$ on the unit sphere, 
the radius vector, the tangent to the curve, and the binormal 
form an orthonormal triad which we call the Darboux triad.}
say $\balpha_L, \btau_L, \bnu_L$ for the charged leptons at any 
conveniently chosen fixed scale, say here $\mu = m_{\tau}$:
\begin{eqnarray}
\balpha_L & = & (-0.90435, 0.31923, 0.28329) \nonumber \\
\btau_L & = & (0.05053, 0.73916, -0.67163) \nonumber \\
\bnu_L & = & (-0.42380, -0.59308, -0.68458)
\label{DbtriadL}
\end{eqnarray}
as well as the rotation angle from the scale $\mu = m_{\tau}$ 
to the scale $\mu = m_{\mu}$, namely:
\begin{equation}
\cos \omega_L = \bmu \cdot \btau_L = - 0.95717, \ \ \ 
\sin \omega_L = \bmu \cdot \bnu_L = - 0.28951.
\label{cosomegaL}
\end{equation}

As explained in \cite{strongCP,atof2cps}, one can then express 
the chirally transformed complex state vector of each of the 3 
charged leptons in terms of the Darboux triad (\ref{DbtriadL}) 
as:
\begin{eqnarray}
\tilde{\btau} & = & \balpha(\mu = m_{\tau}), \nonumber \\
\tilde{\bmu} & = & \cos \omega_L \btau_L 
   + \sin \omega_L \bnu_L e^{i\alpha'}, \nonumber \\
\tilde{\be} & = & - \sin \omega_L \btau_L
   + \cos \omega_L \bnu_L e^{i\alpha'},
\label{statevecscL}
\end{eqnarray}
for $\alpha'$ as given in (\ref{alpha'}).  These formulae are 
the exact parallel of equation (16) in \cite{tfsm} for quarks 
apart from the correction of a misprint on the sign of $\omega$. 
 
A similar set of formulae holds for the neutrinos $N$.  First 
the Darboux triad:
\begin{eqnarray}
\balpha_N & = & (-0.62012, 0.07942, 0.78048) \nonumber \\
\btau_N & = & (-0.76460, 0.16153, -0.62394) \nonumber \\
\bnu_N & = & (-0.17562, -0.98367, -0.03944).
\label{DarbouxN}
\end{eqnarray}
Secondly, the rotation angle from the scale $\mu = m_{\nu_3}$ 
to the scale $\mu = m_{\nu_2}$:
\begin{equation}
\cos \omega_N = \bnu_2 \cdot \btau_N = - 0.99753, \ \ \ 
\sin \omega_N = \bnu_2 \cdot \bnu_N = - 0.06936.
\label{cosomegaN}
\end{equation}
Lastly, the transformed vectors:
\begin{eqnarray}
\tilde{\bnu}_3 & = & \balpha(\mu = m_{\nu_3}), \nonumber \\
\tilde{\bnu}_2 & = & \cos \omega_N \btau_N
   + \sin \omega_N \bnu_N e^{i\alpha'}, \nonumber \\
\tilde{\bnu}_1 & = & - \sin \omega_N \btau_N
   + \cos \omega_N \bnu_N e^{i\alpha'}
\label{statevecscN}
\end{eqnarray}

Together then with the formulae for the charged leptons, one 
has the chirally transformed PMNS matrix as:
\begin{equation}
U_{PMNS} = \left( \begin{array}{ccc} 
           {\tilde{\bnu_1}}^\dagger \tilde{\be}
         & {\tilde{\bnu_2}}^\dagger \tilde{\be}
         & {\tilde{\bnu_3}}^\dagger \tilde{\be}\\
           {\tilde{\bnu_1}}^\dagger \tilde{\bmu}
         & {\tilde{\bnu_2}}^\dagger \tilde{\bmu}
         & {\tilde{\bnu_3}}^\dagger \tilde{\bmu}\\
           {\tilde{\bnu_1}}^\dagger \tilde{\btau}
         & {\tilde{\bnu_2}}^\dagger \tilde{\btau}
         & {\tilde{\bnu_3}}^\dagger \tilde{\btau}
           \end{array} \right),
\label{PMNScomplex}
\end{equation}
where we have followed the usual convention where, for example,
\begin{equation}
\bnu_e = U_{e1} \bnu_1 + U_{e2} \bnu_2 + U_{e3} \bnu_3.
\label{nue}
\end{equation}
This definition (\ref{PMNScomplex}), convenient for FSM, of the 
mixing matrix $U_{PMNS}$ elements as scalar products of state 
vectors in generation space is equivalent to any other definition 
in common use in the current literature.

From this, the Jarlskog invariant can be evaluated, as for 
example:
\begin{equation}
J' = \Im [({\tilde{\bnu_2}}^\dagger \tilde{\bmu})
          ({\tilde{\bnu_3}}^\dagger \tilde{\btau})
          ({\tilde{\bnu_3}}^\dagger \tilde{\bmu})^*
          ({\tilde{\bnu_2}}^\dagger \tilde{\btau})^*].
\label{J'expl}
\end{equation}
In other words, once given the value of $\alpha'$, one can 
calculate explicitly $J'$ to compare with experiment.

We next recall from (\ref{alpha'}) that $\alpha'$ is made up 
of 2 parts.  One, $\theta'_I$, comes in the action from flavour 
instantons, which is only known at present to have a value 
between $0$ and $2 \pi$ and so is naturally of order unity.  
The other $\theta'_C$ comes about as a result of the FSM's 
solution to the strong CP problem and will be there whether or
not flavour instantons exist.  Further, an estimate for 
$\theta'_C$ should already be deducible from the fit to data 
in \cite{tfsm}.  It would be interesting therefore to ignore 
for the moment the $\theta'_I$ term and see what the known 
$\theta'_C$ term will give by itself to the Jarlskog invariant 
$J'$ for leptons, and how it measures up to experiment.

To work out what $\theta'_C$ is, we recall that in \cite{tfsm}
both the up- and down-type quark zero modes were subjected to 
the chiral transformation (\ref{chiraltrans}) with $\alpha$ 
taking the same value $-\alpha_0$ in (\ref{alpha0}).  This is 
appropriate in the context of the flavour theory since they 
are the components of a flavour doublet.  In consequence, each 
quark state affected by this chiral transformation will give 
to the measure of Feynman path integrals an extra factor with  
exponent:
\begin{equation}
- \frac{(-\alpha_0)}{16 \pi^2} \tr_F (G^{\mu \nu} G^*_{\mu \nu}),
\label{extratermFq}
\end{equation}
with only half the value expected from (\ref{extraterm}) because 
in the flavour theory only the left-, but not the right-handed, 
component of $\psi$ couples to the flavour gauge field.  Quarks, 
however, occur in 3 colours, giving thus in total:
\begin{equation}
\theta'_C = - 3 \alpha_0.
\label{theta'C}
\end{equation} 
Again, both the existence of the term (\ref{extratermFq}) and 
its multiplication by factor 3 in (\ref{theta'C}) the deduction 
of which may seem here a little heuristic, will be given a more 
formal derivation in Appendix A.

Putting into (\ref{J'expl}) and the numbers in (\ref{alpha0}) 
and (\ref{theta'C}) above, one obtains:
\begin{equation}
J'_{\rm est} = - 0.0119, 
\label{J'est}
\end{equation}
a measurable quantity which can be confronted with experiment.

The present experimental situation of CP-violation for leptons 
is still rather fluid, with quite different numbers given by 
different analyses.  We cite only the following as examples,
the best fits:
\begin{eqnarray}
J'_{\rm best} & = & - 0.019\ \   \cite{pdg}; 
   \nonumber \\
J'_{\rm best} & = & - 0.0089\ \  \cite{nufit},
\label{J'bestfits}
\end{eqnarray}
and the $1 \sigma$ ranges:
\begin{eqnarray}
J' & = & -0.0304^{+0.0107}_{-0.0025} \ \
   \cite{pdglive} \nonumber \\
J' & = & -0.0086^{+0.0144}_{-0.0214} \ \   
   \cite{nufit} \nonumber \\
J' & = & -0.0084^{+0.0126}_{-0.0117} \ \ 
   \cite{Valle},
\label{J'range}
\end{eqnarray}
inferred respectively from the $1 \sigma$ range of $\delta'_{CP}$:
\begin{eqnarray}
\delta'_{CP} & = & 1.36^{+0.20}_{-0.16} \ \pi,
  \ \  \cite{pdglive} \nonumber \\
\delta'_{CP} & = & 1.08^{+0.28}_{-0.14} \ \pi,
  \ \  \cite{nufit} \nonumber \\
\delta'_{CP} & = & 1.08^{+0.13}_{-0.12} \ \pi,
  \ \  \cite{Valle},
\label{deltarange}
\end{eqnarray}
via the following formula valid in the standard parametrization 
\cite{ChauKeung} in which $\delta'_{CP}$ is defined:
\begin{equation}
J' =  \sin{\delta'_{CP}} J'_{max},
\label{J'fromdel'}
\end{equation}
where
\begin{equation}
J'_{max} = \sin \theta_{12} \sin \theta_{13} \sin \theta_{23}
\cos \theta_{12} \cos^2 \theta_{13} \cos \theta_{23}
\label{J'max}
\end{equation} 
using the values:
\begin{eqnarray}
J'_{max} &=& 0.03359 \ \ \cite{pdg} \nonumber\\ 
J'_{max} &=& 0.0332 \ \ \cite{nufit}.
\label{J'maxvalues}
\end{eqnarray}

We prefer to work with $J'$ rather than $\delta'_{CP}$ because $J'$ is 
parametrization independent, and therefore physically more meaningful 
and easier to calculate. Besides, being a periodic variable  $\delta'_{CP}$
does not contain information about the actual magnitude of CP-violation, 
as $J'$ does.
But one could easily work it the other way 
round.  The fitted values of $\theta_{12}, \theta_{13}, \theta_{23}$ 
in Table 2 of \cite{tfsm} give:
\begin{equation}
J'_{max} = 0.0339,
\label{J'max0}
\end{equation}
which is actually closer to experiment (\ref{J'maxvalues}) than were 
the individual angles, and when applied to (\ref{J'fromdel'}) turns 
(\ref{J'est}) above into an estimate for $\delta'_{CP}$ instead as:
\begin{equation}
\delta_{PC}^{' \rm est} = 1.114 \pi.
\label{del'est}
\end{equation}
This can then be compared directly to (\ref{deltarange}).

\begin{figure}[h]
\centering
\includegraphics[scale=0.325]{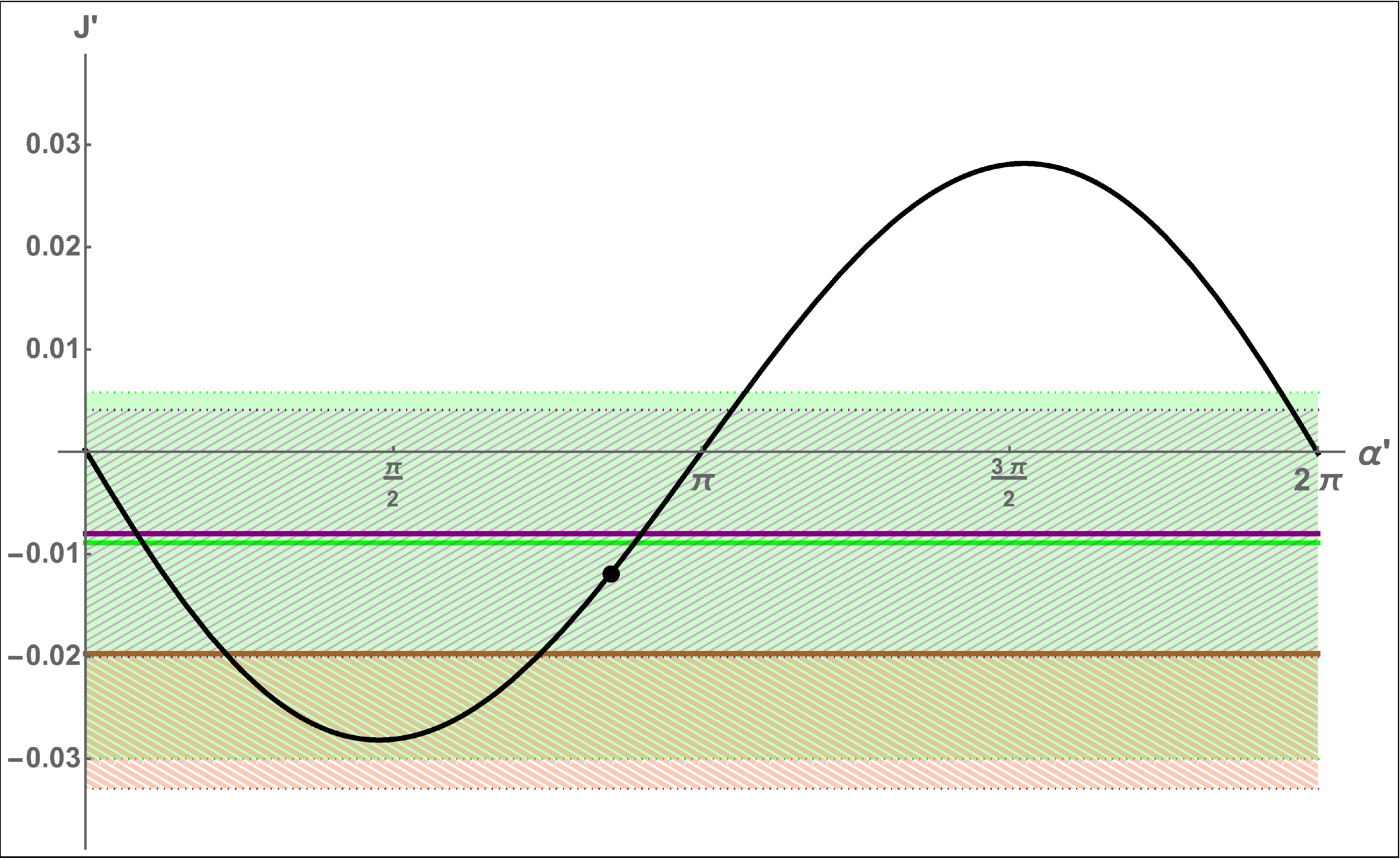}
\vspace*{1cm}
\caption{Comparing FSM estimates to experiment.  Curve shows the
predicted values for $J'$ for various values of $\alpha'$ about 
the preferred value (dot) at $\alpha' = 3 \alpha_0$ suggested by 
\cite{cpslash}.  The lines represent best fit value of $J'$ given 
by respectively \cite{pdg} (brown), \cite{nufit} (green), and 
\cite{Valle} (mauve), while the shaded areas, the 1$\sigma$ range 
of $J'$ inferred from \cite{pdglive} (orange hatch), \cite{nufit} 
(green), and \cite{Valle} (mauve hatch).}  
\label{varialpha'}
\end{figure}

In any case, comparing with the above experimental numbers, one sees 
that (\ref{J'est}) 
\begin{itemize}
\item {\bf (R1)} has the right order of magnitude,
\item {\bf (R2)} is of the right sign, 
\item {\bf (R3)} lies in the range favoured by present experiment,
\end{itemize}
as is depicted in Figure \ref{varialpha'}.

That such is attained is no accident.  From (\ref{J'fromdel'}),
one sees that the order of magnitude of $J'$ is given generically by 
$J'_{max}$ (\ref{J'max}).
Since in \cite{tfsm}, one claimed to have obtained 
more or less the mixing angles right, then ${\bf (R1)}$ should 
follow.  Next, the sign of $\theta'_C$ is necessariy opposite 
to that of $\theta_I$ so as to cure the strong CP problem.  It 
follows then from earlier discussions that $J'$ and $J$ should 
have different signs, namely $J' < 0$, since $J > 0$, that is 
${\bf (R2)}$, at least in the range of values considered.  Then 
given ${\bf (R1)}$, ${\bf (R2)}$ and
the wide range of values of $J'$ still allowed by experiment, 
it will not be hard with a bit of luck to get ${\bf (R3)}$ as well.  
In other words, if for some reason $\theta'_I$ from flavour 
instantons just happens to vanish or, more plausibly, if it is 
nonvanishing but is cancelled by some mechanism other than the 
chiral transformation (\ref{chiraltrans'}) on leptons, then this 
estimate (\ref{J'est}) would be elevated to a prediction of the 
FSM which has come close to hitting the mark.

Interestingly, working concurrently on a hidden sector that the 
FSM also predicts \cite{cpslash}, we have indeed chanced upon
a reason for $\theta'_I$ to be cancelled by another mechanism.
Briefly, it turns out that there is a strong CP problem also 
in the hidden sector, to cure which---in order, for example 
to avoid the universe getting too much dark matter---the 
theta-angle term with $\theta'_I$ as coefficient would have to 
be cancelled by a chiral transformation on some hidden sector 
fermions called $F$.  If so, it leaves in (\ref{alpha'}) above 
only $\theta'_C$ to be cancelled by a chiral transformation on 
leptons and leads then to (\ref{J'est}) as an actual prediction 
of the FSM for the value of the Jarlskog invariant $J'$ in the 
PMNS matrix.

If true, then that $J'$ in (\ref{J'est}) should come at all 
close to experiment would seem highly nontrivial.  It was a 
very long shot, starting in \cite{tfsm} from experiment in hadron 
physics, going through the heart of the FSM scheme all the way to 
topology in the measure of Feynman path integrals and then all 
the way back, again through the very heart of the FSM including 
even the predicted hidden sector, to connect at the end with 
experiments on neutrinos.  And it seems to have landed roughly 
in the right area.  Besides, for the FSM, its validity would serve 
as the first ever check of a basic tenet of the FSM take on CP, 
namely that the CP-violating phase $\delta_{CP}$ in mixing matrices 
has a topological origin.  True, \cite{tfsm} has already shown, 
based on this tenet, that the $\delta_{CP}$ phase in the CKM matrix 
would follow from an instanton term in the action, but there was 
no independent information on the size, namely $\theta_I$, of that 
term which was supposedly cancelled by chiral transformations to 
yield the said phase.  Here, however, the value of the coefficient 
$\theta'_C$ of the topological term is known, and it is explicitly 
the cancellation of this term via chiral transformations that has 
led to $\delta'_{CP}$ or $J'$, leaving little of the logical chain 
in the shade.  Furthermore, it would serve also as a check on the 
form of the FSM mass matrix (\ref{mfact}) with its rotating zero 
mode, without which, of course, the whole FSM tenet on CP would not 
have existed.  Hence, if upheld, this result would be significant, 
both by itself and as a check on the FSM.

It would be appropriate, however, to end this section with a 
couple of cautionary notes:

\begin{itemize}

\item {\bf [C1]} The estimate of $J'$ in (\ref{J'est}) is but a 
crude one, meant only as exploratory not definitive, as it was 
based on a numerical fit with plenty of room for improvement.  

In particular, the value of $\alpha_0$ given in (\ref{alpha0}) is 
constrained in \cite{tfsm} mainly by the experimental values of 
(i) $J$, and (ii) the corner elements of the CKM matrix $|V_{td}|, 
|V_{ub}|$ but, as can be seen in Table 2 there, these quantities 
were actually not that well fitted.  Figure \ref{varialpha'} shows 
how the calculated $J'$ varies with changing $\alpha'$, and it is 
seen that for a fairly wide range of deviations of $\alpha'$ from 
the fitted value $3 \alpha_0$, the predicted value of $J'$ would 
still be experimentally acceptable at present.
 
Of more concern is that $J'$ in (\ref{J'est}) is calculated with 
the lepton state vectors obtained from the fit \cite{tfsm}.  That 
fit, however, was done under the assumption that there was no CP 
phase in the PMNS matrix.  However, now that we have made the 
state vectors complex by chiral transformations, namely into
(\ref{statevecscL}) and (\ref{statevecscN}), the output values 
of the fitted quantities would change and may not agree with data 
as well as they did in the original fit.  This has been checked, 
and indeed, in some cases, the deviations from the original fit 
are quite appreciable.  Hence, the state vectors (\ref{statevecs}) 
we started with may not be as good approximations as the original 
fit indicates.  Instead of the analysis presented above one could 
proceed more consistently, as was done in \cite{tfsm} for quarks,
and redo that whole fit taking account, right from the start, of 
the theta-angle terms for flavour.  This we have not done for lack 
of courage, and also for fear that it may be premature since the 
present data on $J'$ may not yet be sufficiently precise for such 
a refit.  

\item {\bf [C2]}  The estimate (\ref{J'est}) from $\theta'_C$ alone, 
we recall, was taken to be an actual prediction for the value of 
$J'$ by virtue of the conclusion from another paper \cite{cpslash} 
that the $\theta'_I$ term from instantons is to be cancelled by a 
chiral trasnformation on some fermions called F in the hidden sector 
predicted by the FSM \cite{cfsm}.   Although this would seem to be 
the result if one attempts to follow the FSM scheme as far as it 
would lead, many might find it too fanciful, given that even the 
possible existence of the predicted hidden sector cannot as yet be 
subjected to direct experimental test.  There have indeed been some 
attempts to probe the hidden sector's existence \cite{zmixed,fsmanom} 
but the evidence obtained, not surprisingly, is still very flimsy. 
If one insists then, for this reason, on restricting consideration 
to only particles one knows, that is in FSM language, to the standard 
sector, both the theta-angle terms with respective coefficients 
$\theta'_I$ and $\theta'_C$ will have to be cancelled by a chiral 
transformation on leptons.  This is the same as saying that $\alpha'$
in Figure \ref{varialpha'} can take any value from 0 to $2 \pi$.  In 
that case, both ${\bf (R2)}$ and ${\bf (R3)}$ will be lost but, 
except when $\alpha'=0\ {\rm or}\ \pi$,
the $J'$ which results will still be 
of the right 
order of magnitude $10^{-2}$ seen in experiment ${\bf (R1)}$.
 
But even this last assertion is nontrivial.  One recalls \cite{tfsm} 
that also in the colour theory a theta-angle term with $\theta_I$ of 
order unity gives in FSM a Jarlskog invariant $J$ in the CKM matrix 
of the size observed in experiment.  But $J$ is of order $10^{-5}$ 
while $J'$ here is of order $10^{-2}$.  Qualitatively, this difference 
can be understood in the FSM as follows.  CP-violation in the mixing 
matrices comes about in the FSM by virtue of the changing orientation 
(rotation) of the mass matrix, and hence also of the state vectors 
with changing scales.  A look at Figure 3 of \cite{tfsm} shows that 
the rotation is much slower at the higher mass scales of the quarks 
than at the lower mass scales of the leptons.  This in turn, in the 
FSM, is explained, first by a fixed point at $\mu = \infty$ 
and secondly, by a vacuum transition point at $\mu \sim 17$ MeV.  
From this it would follow that the CP-violation as measured by the 
Jarlskog invariant is smaller for quarks than for leptons, as observed.  
However, to see that $J$ and $J'$ should differ by 3 orders of 
magnitude, we have had to perform the actual calculation in detail as 
prescribed above by the FSM, offering thereby a check on the model's
internal consistency.

\end{itemize}

\section{Unified treatment of all known CP physics?} 

We recall that in both the colour and the flavour cases, it 
was shown that theta-angle terms in Feynman integrals can be 
cancelled by appropriate chiral transformations on some fermion 
states with zero mass eigenvalues inherent in the FSM and 
turned into a CP-violating phase in the mixing matrices.  
Now that one has further checked that the resulting Jarlskog 
invariants $J$ and $J'$, measuring the amount of CP-violation, 
are of the same order of magnitude as those experimentally 
observed, the possibility then exists that all CP-violation 
effects that have been seen so far do indeed arise in the way 
that the FSM prescribes.  If that is the case, a rather 
pleasing picture results.
           
Let us end then with a resum\'e of the resultant situation in 
the FSM relative to that in the standard model.  

In the SM, our common conception seems to be, as regards CP 
and its violation, that there is a difference in substance 
between the colour and flavour sectors, and hence between 
quarks which carry both colour and flavour and leptons which 
carry only flavour.  In the colour sector, there is, firstly, 
a strong CP problem in the form of a theta-angle term 
(\ref{thetatermC}) in the action which is topological in 
origin (instantons), where $\theta_I$ can take any value between 
0 and $2 \pi$, and hence is naturally of order unity.  This 
can lead in principle to CP-violation in strong interaction 
effects, such as the neutron electric dipole moment, which are 
$\gtrsim 9$ orders in magnitude higher than that allowed by 
experiment \cite{neutrondipole}.  We are thus saddled with the 
question why such huge CP-violations are not observed.  The 
most widely favoured answer for this is a new $u(1)$ dynamics 
\cite{PecceiQ} which leads in turn to new particles called 
axions \cite{WeinWil}, but such particles, in spite of intense 
searches for decades by experiment, have not yet been found.  
Secondly, still in the colour sector, the standard model 
predicts that there can be CP-violation in the CKM matrix in 
the form of a KM phase \cite{KM} or equivalently a Jarlskog 
invariant \cite{Jarlskog}.  This has been measured \cite{pdg}, 
thus confirming the SM prediction of its possible existence, 
but the SM gives no estimate for its size nor any idea of its 
physical origin.  And in the SM, no connection is known to 
exist between the strong CP problem as characterised by the 
theta-angle term and the weak CP-problem as characterized 
by the KM phase.  Hence the strong CP angle $\theta_I$ and the 
KM phase $\delta_{CP}$ count as two independent parameters.  
Next, in the flavour sector, although a theta-angle term 
(\ref{thetatermF}) should also appear in the action for 
basically the same reason as in colour, the idea used to be 
that it led to no physical effects (for reasons that we now 
find inadequate, as argued in Appendix B).  In any case, 
this still leaves us with the possibility of a CP-violating 
phase in the PMNS matrix similar to the KM phase in the CKM.  
But again, for the size and physical origin of this phase or 
of the corresponding Jarlskog invariant, the SM gives no 
indication.  This phase has thus again to be taken as another 
empirical parameter.   
  
In the FSM, on the other hand, in the treatment as described 
above:
\begin{itemize}
\item CP-violations for quarks and leptons are put on the same 
footing.
\item They both have a topological origin (instantons) via
theta-angle terms in the action, one in the colour theory 
($\theta_I$) and one in the flavour theory ($\theta'_I$).
\item Although both $\theta_I$ and $\theta'_I$ are naturally 
of order unity with potentially huge CP-violating effects much 
beyond those observed in experiment, both these theta-angle 
terms can be cancelled in Feynman path integrals, at any fixed 
scale $\mu$, by appropriate chiral transformations on some 
quark and lepton states with zero mass eigenvalues inherent 
in the theory, avoiding thus all huge CP-violations.  This 
solves in particular the ``strong CP problem'' (with no new 
dynamics or axions introduced).
\item In either case, the chiral transformation performed 
can be understood physically as a choice of the relative 
phase between the left- and right-handed components in the 
definition of CP for the quark and/or lepton fields of zero 
mass eigenvalues, for which no {\it a priori} choice exists.  
\item But the state vectors of these quark and lepton states 
with zero mass eigenvalues, on which the above chiral 
transformations act, depend on scale $\mu$ by virtue of the 
RGE derived from framon loops (same RGE as that which gave 
previously in the FSM the mass and mixing patterns of quarks 
and leptons \cite{tfsm}).  This $\mu$-dependence is what 
gives rise to the CP-violating phases $\delta_{CP}$ and 
$\delta'_{CP}$ in respectively the CKM and PMNS matrix.  And 
the ensuing CP-violating effects from these phases, whether 
in quarks or leptons, by virtue of being loop effects, are 
bound to be perturbatively small.
\item Thus, $\theta_I$ and $\delta_{CP}$, being inter-related, 
count together as only one parameter.  So do $\theta'_I$ and 
$\delta'_{CP}$ for leptons for the same reason.  
\item A $\theta_I$ ($\theta'_I$) of order unity is found to lead 
automatically to Jarlskog invariant of the order of magnitude 
$10^{-5}$ ($10^{-2}$) seen in experiment.  [Indeed, if we take 
further account of results in the hidden sector, we have even 
a prediction for the Jarlskog invariant $J'$ for leptons which
lies roughly in the range favoured by present experiment.]
\end{itemize}

Thus, in brief, in both the colour and flavour case, the 
potential huge violations from theta-angle terms are reduced 
to just a CP-violating phase in the mixing matrix which are 
indeed the only CP-violating effects so far observed.  Hence, 
if the arguments put forth above, both theoretical and 
numerical, can survive the scrutiny of the community, one 
might have found here a much neater and more unified approach 
to CP physics that we have ever had before. 

\vspace*{1cm}

\noindent{\bf\large Appendix A: More formal derivation of certain 
stated results}

\vspace*{7mm}

In the notation used in this article,
\begin{eqnarray}
G^{\mu \nu} & = & G_i^{\mu \nu} t_i, \nonumber\\ 
H^{\mu \nu} & = & H_K^{\mu \nu} t_K,
\label{GHdef}
\end{eqnarray}
where repeated indices are to be summed, and $t_i, i = 1, 2, 3$ and 
$t_K, K = 1, ..., 8$ are generators of respectively the flavour 
$su(2)$ and colour $su(3)$ symmetries normalized such that: 
\begin{eqnarray}
\tr_F (t_i t_j) & = & \half \delta_{ij}; \nonumber\\ 
\tr_C (t_K t_L) & = & \half \delta_{KL},
\label{trFtrC}
\end{eqnarray}
and
\begin{eqnarray}
G^*_{\mu \nu} & = & \half \epsilon_{\mu \nu \rho \sigma} G^{\rho \sigma}, \nonumber
 \\
H^*_{\mu \nu} & = & \half \epsilon_{\mu \nu \rho \sigma} H^{\rho \sigma}.
\label{GHstardef}
\end{eqnarray}

It follows then that:
\begin{eqnarray}
\tr_F (G^{\mu \nu} G^*_{\mu \nu}) & = & \quart \epsilon^{\mu \nu \rho \sigma} 
                  G_{i\mu \nu} G_{i\rho \sigma}; \\
\tr_C (H^{\mu \nu} H^*_{\mu \nu}) & = & \quart \epsilon^{\mu \nu \rho \sigma} 
                  H_{K\mu \nu} H_{K\rho \sigma}.
\end{eqnarray}
and the theta-angle terms become:
\begin{eqnarray}
{\cal L}_{\theta'} & 
   = & - \frac{\theta'_I}{16 \pi^2} \tr_F (G^{\mu \nu} G^*_{\mu \nu}) \\
{\cal L}_\theta & 
   = & - \frac{\theta_I}{16 \pi^2} \tr_C (H^{\mu \nu} H^*_{\mu \nu}),
\end{eqnarray}
where the latter accords with (23.6.6) in \cite{Weinbergqft} apart 
from the change in notation $F \rightarrow H$, and $\theta \rightarrow \theta_I$.

As to the difficult question first answered by Fujikawa 
\cite{Fujikawa} how the measure of Feynman path integrals transforms under 
a chiral transformation of a fermion field interacting with a gauge 
field:
\begin{equation}
\psi \longrightarrow (\exp i \alpha \gamma_5) \psi,
\label{chiraltransa}
\end{equation}
we are happy to find in Section 2.2 of \cite{Weinbergqft} an  
understandable treatment which gives, for the special case we need 
(when $\alpha$ is $x$-independent and $t = 1$ in the notation 
therein):
\begin{equation}
[d \psi][d \bar{\psi}] \rightarrow 
   \exp \left\{ i \alpha \int d^4 x {\cal A}(x) \right\} 
   [d \psi][d \bar{\psi}],
\label{meastrans}
\end{equation}
where the anomaly function ${\cal A}$, when appropriately regularized, 
takes the following form:
\begin{equation}
{\cal A}(x) = - \frac{i}{16 \pi^2} \Tr \left\{\frac{1}{16} \gamma_5 
                ([(D_x)^\mu, (D_x)^\nu] [\gamma_\mu, \gamma_\nu])^2
                \right\},
\label{calA1}
\end{equation}
where:
\begin{equation}
\Tr = \tr_D \tr_{IS},
\label{Tr}
\end{equation}
with $\tr_D$ being the trace over Dirac indices and $\tr_{IS}$ over 
internal symmetry indices.  Hence, using
\begin{equation}
\tr_D \{ \gamma_5 [\gamma_\mu, \gamma_\nu] [\gamma_\rho, \gamma_\sigma] \}
   = 16 i \epsilon_{\mu \nu \rho \sigma},
\label{trgammas}
\end{equation}
one gets:
\begin{equation}
{\cal A}(x) = \frac{1}{16 \pi^2} \tr_{IS} \{ \epsilon_{\mu \nu \rho \sigma}
              [(D_x)^\mu, (D_x)^\nu][(D_x)^\rho, (D_x)^\sigma] \}.
\label{calA2}
\end{equation}

Specializing now to colour $su(3)$, one gets:
\begin{equation}
{\cal A}(x) = - \frac{(2)}{16 \pi^2} \tr_C \{ H^{\mu \nu} H^*_{\mu \nu}\},
\label{calAC}
\end{equation} 
which accords with (\ref{extraterm}) in the text.  The factor of 2 can 
conveniently be thought of as coming 1 each from the left- and right- 
handed quarks both of which are coupled to the colour gauge field.  Whilst 
specializing to flavour $su(2)$ one gets,
\begin{equation}
{\cal A}(x) = - \frac{(2)}{16 \pi^2} \tr_F \{ G^{\mu \nu} G^*_{\mu \nu}\},
\label{calAF}
\end{equation}
which accords with (\ref{alpha'}) in the text apart from the factor 2.  
This difference is due, as argued in the text, just to the fact that 
the flavour theory is chiral where only left-handed quarks and leptons 
are coupled to the flavour gauge field, a fact that we have not yet 
learned to incorporate in the present more formal treatment copied from 
\cite{Weinbergqft} meant only for nonchiral gauge fields. 

What actually interests us in this paper, however, is the theory where 
the gauge symmetry is $su(2) \times su(3)$ and where the quark field 
undergoing chiral transformation is a doublet in flavour $su(2)$ and a 
triplet in colour $su(3)$.  In this case,
\begin{equation}
(D_x)_\mu = \partial_\mu - i  B_{i \mu} t_i - i  C_{K \mu} t_K,
\label{Dxmu}
\end{equation}
and
\begin{equation}
\Tr = \tr_D \tr_F \tr_C,
\label{Tr}
\end{equation}
with $\tr_D, \tr_F, \tr_C$ denoting respectively the trace over Dirac, 
flavour, and colour indices.  Since $t_i$ and $t_K$ commute,
\begin{equation}
[(D_x)^\mu, (D_x)^\nu] = -i G_i^{\mu \nu}(x) t_i -i H_K^{\mu \nu}(x) t_K.
\label{Dcom}
\end{equation} 
Then, using again (\ref{trgammas}) one gets:
\begin{equation}
{\cal A}(x) = - \frac{1}{16 \pi^2} \epsilon_{\mu \nu \rho \sigma}
   \tr_F \tr_C \{ (G_i^{\mu \nu}(x) t_i  + H_K^{\mu \nu}(x) t_K) 
   (G_j^{\rho \sigma}(x) t_j  + H_L^{\rho \sigma}(x) t_L) \}.
\label{calAFC}
\end{equation}
The cross terms vanish under $\tr_F$ or $\tr_C$ leaving:
\begin{equation}
{\cal A}(x) = - \frac{1}{16 \pi^2} \epsilon_{\mu \nu \rho \sigma}
 [\tr_F \{(G_i^{\mu \nu}(x) t_i) (G_j^{\rho \sigma}(x) t_j)\} \tr_C (1) +  
  \tr_C \{(H_K^{\mu \nu}(x) t_K) (H_L^{\rho \sigma}(x) t_L)\} \tr_F (1)]
\label{calAFC2}
\end{equation}
or, using (\ref{GHstardef}), and $\tr_F(1) = 2$ and $\tr_C(1) = 3$:
\begin{equation}
{\cal A}(x) = - \frac{1}{16 \pi^2} (2) [3 \tr_F (G^{\mu \nu} G^*_{\mu \nu})
                 + 2 \tr_C (H^{\mu \nu} H^*_{\mu \nu})].
\label{calAFC3}
\end{equation}

In other words, under the chiral transformation (\ref{chiraltransa}) of 
the quark field in the $su(2) \times su(3)$ theory, one gets an effective
extra term in the Lagrangian density of the form:
\begin{equation}
- \frac{(6 \alpha)}{16 \pi^2} \tr_F (G^{\mu \nu} G^*_{\mu \nu})
     - \frac{(4 \alpha)}{16 \pi^2} \tr_C (H^{\mu \nu} H^*_{\mu \nu})].
\label{extratermFC}
\end{equation}
We note then the following:
\begin{itemize}
\item The factor in front of the $\tr_C (H H^*)$ term is $4 \alpha$, which 
is twice that in the pure colour $su(3)$ theory (\ref{calAC}), receiving 
as it does a contribution of $2 \alpha$ each from the up-type and down-type 
quark, in accord with the argument in the text.  That the up- and down-type 
quarks should give the same contribution comes from the fact that they are 
members of the same flavour doublet transforming under (\ref{chiraltransa}),
a fact used already in the fit \cite{tfsm} but not elucidated there.  To 
cancel the theta-angle term from instantons (\ref{thetatermC}) in the text,
one needs then $4 \alpha = - \theta_I$, as stipulated in (\ref{thetaI}) in 
the text. 
\item The chiral transformation on the quark field suggested for cancelling 
the $\theta_I$ term in colour does indeed produce in (\ref{extratermFC}) also 
a $\tr_F (G G^*)$ term, as asserted in the text.  In (\ref{extratermFC}) this 
comes with a factor $6 \alpha$ which, according to the rules of accounting 
stated after (\ref{calAF}) above, should be halved because the flavour 
theory is chiral.  This leaves then a factor of $3 \alpha$ because quarks 
come in 3 colours, in accord with the discussion leading to (\ref{theta'C}) 
in the text.
\end{itemize}  

\vspace*{1cm}

\noindent{\bf\large Appendix B: Physical effects from a theta-angle 
term in the flavour theory?}

\vspace*{7mm}

Let us then take a closer look at the arguments that supposedly 
led to {\bf [S]}.  We shall proceed cautiously, even pedantically, 
so as to make the logic as clear as we possibly can.  Surprisingly, 
a recent search by us shows no articles devoted to the question, 
only some articles where {\bf [S]} appears as a side remark.  The 
treatments vary, but those we have seen and, we think, understood 
can be paraphrased as follows.  

Subject any flavour doublet fermion (4-component Dirac) field with 
a nonzero mass term, quark or lepton, to the chiral transformation: 

\begin{equation}
\psi \rightarrow (\exp i \alpha' \gamma_5) \psi.
\label{chiraltrans'A}
\end{equation}
This will give to the exponent of the measure in the Feynman path 
integral an extra term \cite{Weinbergqft,Fujikawa} in the Lagrangian 
density, the flavour theory being chiral:
\begin{equation}
- \frac{\alpha'}{16 \pi^2} \Tr (G^{\mu \nu} G^*_{\mu \nu}).
\label{extraterm'}
\end{equation} 
Hence by choosing $\alpha'$, summed over all the fermions, to be 
$- \theta'_I$, the theta-angle term (\ref{thetatermF}) is cancelled 
in Feynman path integrals.  

However, this would leave the mass term of $\psi$ complex.  To 
see this, we write as usual:
\begin{equation}
\psi = \psi_L + \psi_R,
\label{psiLR}
\end{equation}
with
\begin{equation}
\psi_L = \half (1 + \gamma_5) \psi, \ \ 
\psi_R = \half (1 - \gamma_5) \psi.
\label{psiLRdef}
\end{equation}

Under the chiral transformation (\ref{chiraltrans'A}) above,
\begin{equation}
\psi_L \rightarrow e^{i \alpha'} \psi_L, \ \ 
\psi_R \rightarrow e^{-i \alpha'} \psi_R.
\label{psiLRct'}
\end{equation}
The mass term for $\psi$, with the mass parameter $m$ real to 
start with, will then transform under the chiral transformation 
(\ref{chiraltrans'A}) as:
\begin{eqnarray}
m(\bar{\psi} \psi) & = & m[(\bar{\psi}_L)(\psi_R) 
                       + (\bar{\psi}_R)(\psi_L)] \\
& \rightarrow & m \left[ (\bar{\psi}_L e^{-i \alpha'})
     (e^{-i \alpha'}\psi_R) + (\bar{\psi}_R e^{i \alpha'})
     (e^{i \alpha'}\psi_L) \right]\\
& = & m \left[ e^{-2i \alpha'} (\bar{\psi}_L \psi_R) +
               e^{2i \alpha'} (\bar{\psi}_R \psi_L) \right].
\label{mcomplex}
\end{eqnarray}
This is not acceptable since a complex mass term would violate CP.             

However, one can restore reality to the mass value by performing a 
simple phase rotation to the right-handed component of $\psi$, thus:
\begin{equation}
\psi_R \rightarrow (\exp 2i \alpha') \psi_R.
\label{psiRphase}
\end{equation}  
This will not further alter the measure in the Feynman integral or 
affect the cancellation of the theta-angle term since $\psi_R$ is a 
flavour singlet.  One has thus transformed away, as claimed, the 
theta-angle term in the flavour theory, apparently with no other 
undesirable consequences.  Hence, the argument goes, {\bf [S]} 
follows.

We have no objection to the above analysis, which looks all clear 
to us, except for the last sentence, which seems to us to require 
further scrutiny.  The chiral transformation (\ref{chiraltrans'A}) 
on $\psi$ as exhibited in (\ref{psiLRct'}) is followed here by the 
transformation (\ref{psiRphase}), which means in total just a phase 
rotation on $\psi$:
\begin{equation}
\psi \rightarrow (\exp i \alpha') \psi.
\label{psiphase}
\end{equation}

If we make the same phase rotation to all the fermions in the theory, 
which the above argument leading supposedly to {\bf [S]} allows, 
it means just a change in the convention of how we define fermion 
phases.  We obtain then in total:
\begin{equation}
n_f \alpha' \Tr (G^{\mu \nu} G^*_{\mu \nu}).
\label{extraterm'A}
\end{equation} 
with $n_f$ being the total number of fermions, quarks and leptons, in 
the theory.  Choosing $n_f \alpha' = - \theta'_I$, one can cancel the 
theta-angle term (\ref{thetatermF}) in all Feynman path integrals.  
Hence, if we accept the above argument in full and conclude that 
{\bf [S]} then follows, we are saying that the original theory with 
the theta-angle term (\ref{thetatermF}) which explicitly violates CP
to order unity has somehow become CP-conserving merely by changing 
the convention how fermion phases are measured, a convention which 
we are used to regarding as without physical significance.  This is, 
to say the least, counter-intuitive.  Further, even more disturbingly, 
we need not, of course, have chosen $n_f \alpha' = - \theta'_I$.  By 
varying $\alpha'$, we could have transformed the theory instead into 
one with a theta-angle of any value.  In other words, if the argument 
were right, we could, by changing the convention of how we measure 
fermion phases, change the amount of CP-violation which is explicitly 
exhibited.

What then has gone wrong?  In coming to the conclusion {\bf [S]} in
the argument above, one has tacitly assumed that the theory obtained 
after the transformation (\ref{psiphase}) is the same, or at least 
have the same physical content, as that before the transformation is 
performed.  And sure enough, the action is invariant under such a 
transformation.  But that by itself is insufficient to guarantee that 
a quantum theory constructed with that action is itself invariant 
under that same transformation for, in Fujikawa's 
language \cite{Fujikawa}, 
the quantum theory might be anomalous, as it appears to be 
the case here \cite{ABJ}.  That being so, it would be wrong to assert 
as one did that because the theory obtained after the transformation 
(\ref{psiphase}) is CP-conserving, then the original theory before 
the transformation is CP-conserving as well, or that the theta-angle 
term in the original theory has no physical significance.  The two 
theories before and after the transformation are different theories 
altogether, and have thus no reason to have the same CP content.   

This does not mean that different choices of phase conventions have 
any real significance.  For every choice of the phase convention, 
$\theta'_I$ can take any value from $0$ to $2 \pi$, so that we have in 
fact for every choice of phase convention the same family of theories 
parametrized by $\theta'_I$.  In the original choice of phase convention, 
say we have some $\theta'_I \not=0$ as the coefficient of the theta-angle 
term, meaning that we have picked out from that family a specific 
theory corresponding to the parameter $\theta'_I$.  If our theory were 
invariant under phase convention changes, a change in phase convention 
would give us back a theory with a theta-angle term having again the 
coefficient $\theta'_I$, and physics would remain unchanged.  Now however, 
because the (quantum) theory is anomalous, when we perform an actual 
change in the phase convention by the phase rotation (\ref{psiphase}), 
we obtain instead a different theory altogether, namely one with a 
different value for the theta-angle parameter: $\theta'_I+\alpha'$, and 
hence also different physical content under CP.  In other words, that 
the theory obtained after the phase rotation is CP-invariant has no 
reason to imply that the original theory before the rotation is also 
CP-invariant.

It seems to us therefore that {\bf [S]} is a false conclusion to draw from 
the arguments presented, which by themselves do not preclude theta-angle 
terms in the flavour theory from yielding physical effects.

One circumstance which might have contributed to the confusion leading 
to the above false conclusion was the apparent similarity of the argument 
given in the earlier half of this Appendix B to that given in the Preamble for 
the cancellation of the theta-angle term in QCD by a chiral transformation 
on quark states with no mass terms, leading thereby to a solution of the 
strong CP problem.  This similarity, however, is only formal, the two 
arguments being in fact very different in physical content.  In the latter,  
as already explained, the chiral transformation is just a means to specify 
what CP means for the quark states with no mass terms, which has so far 
been left undefined, but the underlying theory is left unchanged.  This 
definition of CP for the quark zero mode is a physical assertion.  It is 
therefore understandable that it could lead to a physical result, namely 
the suppression of the potentially huge CP-violating effects from the 
theta-angle term, hence solving the strong CP problem.  The former, on the 
other hand, is supposedly a transformation of the theory under a change 
of phase convention for fermions, which leaves the action invariant but 
maps the original theory to another because the (quantum) theory is 
anomalous.  What new physics this last fact may lead to has not been made 
clear, but if there is any, it will only go towards refuting {\bf [S]} 
and confirm our earlier assertion.

However, since we have, we claim, already shown in the main text that the 
theta-angle term in flavour will lead in the FSM to a CP-violating phase 
in the PMNS matrix for leptons, which is a measurable physical effect, in 
fact already almost measured in experiment, one has in this result already 
a counter-example to {\bf [S]} which invalidates it, independent of the 
arguments set out in this Appendix.  And since in FSM the flavour theory 
is chiral, the condition required for {\bf [S]} supposedly to hold, the 
appearance of a CP-violating phase for leptons in FSM resulting from the 
theta-angle term (\ref{thetatermF}) serves as a valid counter-example to 
{\bf [S]} irrespective of whether one believes in all the other features 
built into the FSM or not.

\end{document}